\documentclass{revtex4}
\usepackage{amsmath}

\input{tcilatex}

\begin{document}

\preprint{}
\title[ ]{Note on: "Domain wall universe in the Einstein--Born--Infeld
theory" Phys. Lett. B 679 (2009) 160. }
\author{S. Habib Mazharimousavi}
\email{habib.mazhari@emu.edu.tr}
\author{M. Halilsoy}
\email{mustafa.halilsoy@emu.edu.tr}
\affiliation{Department of Physics, Eastern Mediterranean University, G. Magusa, north
Cyprus, Mersin 10, Turkey. }

\begin{abstract}
The interaction between bulk and dynamic domain wall in the presence of a
linear / non-linear electromagnetism make energy density, tension and
pressure on the wall all variables, depending on the wall position. In Ref. 
\cite{1} this fact seems to be ignored.
\end{abstract}

\maketitle

The $(n+1)-$dimensional bulk space time with $Z_{2}$ symmetry can
equivalently be chosen as (i.e. Eq. (4) of Ref. \cite{1})%
\begin{equation}
ds^{2}=-f\left( R\right) dT^{2}+\frac{dR^{2}}{f\left( R\right) }%
+R^{2}d\Omega _{n-1}^{2}
\end{equation}%
in which $d\Omega _{n-1}^{2}$ is the line element on $S^{n-1}.$ The
n-dimensional domain wall (DW) in the FRW form is%
\begin{equation}
ds^{2}=-d\tau ^{2}+a\left( \tau \right) ^{2}d\Omega _{n-1}^{2}
\end{equation}%
with the constraint 
\begin{equation}
f\left( a\right) \dot{T}^{2}-\frac{\dot{a}^{2}}{f\left( a\right) }=1
\end{equation}%
in which a dot implies $\frac{d}{d\tau }.$

The Israel junction condition 
\begin{equation}
\left[ K_{\mu \nu }-g_{\mu \nu }K\right] =-\kappa _{n+1}^{2}S_{\mu \nu }
\end{equation}%
leads to (with $Z_{2}$ symmetry) 
\begin{eqnarray}
-\frac{2\left( n-1\right) }{a}\sqrt{f+\dot{a}^{2}} &=&\kappa
_{n+1}^{2}\left( \rho +\sigma \right) ,\text{ for }\tau \tau \text{ component%
} \\
\frac{2\left( n-2\right) }{a}\sqrt{f+\dot{a}^{2}}+\frac{f^{\prime }+2\ddot{a}%
}{\sqrt{f+\dot{a}^{2}}} &=&\kappa _{n+1}^{2}\left( p-\sigma \right) ,\text{
for }\theta _{i}\theta _{i}\text{ components.}
\end{eqnarray}%
As considered in Ref. \cite{1} the DW energy momentum $S_{\mu }^{\nu }=$diag$%
\left( -\rho -\sigma ,p-\sigma ,...,p-\sigma \right) $ is given by%
\begin{equation}
S^{\mu \nu }=\frac{2}{\sqrt{-g}}\frac{\delta }{\delta g_{\mu \nu }}\int
d^{n}x\sqrt{-g}\left( -\sigma +\mathcal{L}_{m}\right) .
\end{equation}%
in which (Eq. (22) of Ref. \cite{1}) 
\begin{equation}
\mathcal{L}_{m}=\mathcal{L}_{0}+\frac{C}{a^{n-1}}\bar{A}_{\tau }
\end{equation}%
and $C=\pm \frac{q\sqrt{2\left( n-1\right) \left( n-2\right) }}{\kappa
_{n+1}^{2}}.$ By using (7) one finds%
\begin{equation}
S_{\mu \nu }=-2\frac{\delta \mathcal{L}_{DW}}{\delta g^{\mu \nu }}+\mathcal{L%
}_{DW}g_{\mu \nu }
\end{equation}%
for $\mathcal{L}_{DW}=\left( -\sigma +\mathcal{L}_{m}\right) .$ The latter
equation implies (See the Appendix)%
\begin{equation}
S_{\tau }^{\tau }=\frac{2C}{a^{n-1}}\bar{A}_{\tau }+\mathcal{L}_{0}-\sigma ,
\end{equation}%
and 
\begin{equation}
S_{\theta _{i}}^{\theta _{i}}=\mathcal{L}_{0}-\sigma \text{, \ \ }\left(
i=1...n-1\right) .
\end{equation}%
Comparison with the general form of $S_{\mu }^{\nu }$ implies that the
induced electrostatic energy density on the DW is%
\begin{equation}
\rho =-\frac{2C}{a^{n-1}}\bar{A}_{\tau }-\mathcal{L}_{0}
\end{equation}%
while the pressure is 
\begin{equation}
p=\mathcal{L}_{0}.
\end{equation}%
Now, taking into account Eq.s (5) and (6), we get two equations to be
satisfied simultaneously, i.e.%
\begin{equation}
-\frac{2\left( n-1\right) }{a}\sqrt{f+\dot{a}^{2}}=\kappa _{n+1}^{2}\left( -%
\frac{2C}{a^{n-1}}\bar{A}_{\tau }-\mathcal{L}_{0}+\sigma \right)
\end{equation}%
and 
\begin{equation}
\frac{2\left( n-2\right) }{a}\sqrt{f+\dot{a}^{2}}+\frac{f^{\prime }+2\ddot{a}%
}{\sqrt{f+\dot{a}^{2}}}=\kappa _{n+1}^{2}\left( \mathcal{L}_{0}-\sigma
\right) .
\end{equation}%
Herein $\bar{A}_{\tau }$ is given in terms of the bulk potential and metric
function by 
\begin{equation}
\bar{A}_{\tau }=\bar{A}_{T}\frac{\sqrt{f+\dot{a}^{2}}}{f}.
\end{equation}%
The angular part of Israel equation admits%
\begin{equation}
\kappa _{n+1}^{2}\left( \sigma -\mathcal{L}_{0}\right) =-\left( \frac{%
2\left( n-2\right) }{a}\sqrt{f+\dot{a}^{2}}+\frac{f^{\prime }+2\ddot{a}}{%
\sqrt{f+\dot{a}^{2}}}\right) ,
\end{equation}%
which is clearly not a constant. In Ref. \cite{1} the authors consider a new
constant parameter $\chi ^{2}=\kappa _{n+1}^{2}\left( \sigma \ell \right)
^{2}/4\left( n-1\right) ^{2}$ and by setting $\mathcal{L}_{0}=0$ (i.e. zero
pressure) they find an equation of motion for the dynamic domain wall, based
only on Eq. (14), which reads%
\begin{equation}
\dot{a}^{2}+V\left( a\right) =0.
\end{equation}%
Plotting rescaled form of $V\left( a\right) $ for fixed values of $\chi $
(namely $\chi =1.1)$ is the last stage of Ref. \cite{1}. Based on our
argument on the other hand setting $\chi $ to a constant value is equivalent
to setting $\sigma =cons.$ which is obviously in contradiction with the form
of $\sigma $ we found in Eq. (17) above. In other words, choosing $\sigma
=const.$ does not satisfy both of the Israel junction conditions at the same
time.

Unlike this case, if we neglect the interaction between the bulk and domain
wall in the form of Nambu-Goto action, i.e.%
\begin{equation}
S_{DW}=-\sigma _{\circ }\int\limits_{\Sigma }d^{n}x\sqrt{-g}
\end{equation}%
we observe that 
\begin{equation}
S_{\mu \nu }=-2\frac{\delta \mathcal{L}_{DW}}{\delta g^{\mu \nu }}+\mathcal{L%
}_{DW}g_{\mu \nu }=-\sigma _{\circ }g_{\mu \nu }.
\end{equation}%
This means from $S_{\mu }^{\nu }=$diag$\left( -\rho -\sigma ,p-\sigma
,...,p-\sigma \right) =$diag$\left( -\sigma _{\circ },-\sigma _{\circ
},...,-\sigma _{\circ }\right) $ that $-\rho =p=const.$ (which is set to
zero for simplicity). As a result the two Israel junction conditions are
consistent, i.e.%
\begin{eqnarray}
-\frac{2\left( n-1\right) }{a}\sqrt{f+\dot{a}^{2}} &=&\kappa
_{n+1}^{2}\sigma _{\circ } \\
\frac{2\left( n-2\right) }{a}\sqrt{f+\dot{a}^{2}}+\frac{f^{\prime }+2\ddot{a}%
}{\sqrt{f+\dot{a}^{2}}} &=&-\kappa _{n+1}^{2}\sigma _{\circ }.
\end{eqnarray}%
By differentiating (21) one obtains%
\begin{equation}
\frac{f^{\prime }+2\ddot{a}}{\sqrt{f+\dot{a}^{2}}}=\frac{2\sqrt{f+\dot{a}^{2}%
}}{a},
\end{equation}%
which reduces (22) to (21). Therefore these two equations amount to the
single Eq. (21).

Our conclusion to this problem simply implies a more complicated equation of
motion for the dynamic domain wall that emerges from the substitution of
Eq.s (17) into (14) i.e.,%
\begin{equation}
\ddot{a}+\frac{\left( G-1\right) }{a}\dot{a}^{2}+\frac{\left( G-1\right) f}{a%
}+\frac{f^{\prime }}{2}=0,
\end{equation}%
in which 
\begin{equation}
G=\kappa _{n+1}^{2}\left( \frac{C}{a^{n-2}}\bar{A}_{T}\frac{1}{f}\right) .
\end{equation}%
Given the complexities of $f(R)$ and $\bar{A}_{T}$ for the
Einstein-Born-Infeld theory \cite{1}, Eq. (24) is a rather difficult
differential equation to be solved. To give an idea about its structure yet
we resort to the $5-$dimensional cosmological Einstein-Maxwell theory ($n=4$
and $\beta \rightarrow \infty $ limit of Ref. \cite{1}). Solution for $f(R)$
and $\bar{A}_{T}$ are given (from Eq. (12) and (16) of \cite{1} with $\beta
\rightarrow \infty $) by 
\begin{equation}
f\left( R\right) =1+\frac{R^{2}}{\ell ^{2}}-\frac{m^{2}}{R^{2}}+\frac{q^{2}}{%
R^{4}}
\end{equation}%
\begin{equation}
\bar{A}_{T}=\frac{\sqrt{3}}{2}\frac{q}{R^{2}}.
\end{equation}%
Plugging these expressions with (25) into (24) (for $\kappa _{n+1}^{2}=1,$ $%
C=-2\sqrt{3}q$ and $R=a\left( \tau \right) $) plots the $f(R)$ which in turn
determine numerical integrations of (24) for specific parameters. We remark,
that depending on the initial conditions and parameters falling into black
hole or escaping to infinity and any possibility in between those two
extremes are available. We plot, for instance in Fig. 1 the bouncing
property of $a\left( \tau \right) $ with the choice $C<0.$ It should be
remarked that with the choice $C>0$, there is no bounce.

\bigskip \bigskip 

\textbf{APPENDIX:}

To find $S_{\tau }^{\tau }$ and $S_{\theta _{i}}^{\theta _{i}}$ we use the
formula (9), and consider 
\begin{equation}
\mathcal{L}_{DW}=\left( -\sigma +\mathcal{L}_{m}\right) ,
\end{equation}%
in which 
\begin{equation}
\mathcal{L}_{m}=\mathcal{L}_{0}+\frac{C}{a^{n-1}}\bar{A}_{\tau }=\mathcal{L}%
_{0}+\frac{C}{a^{n-1}}\bar{A}^{\mu }g_{\mu \tau }.
\end{equation}%
Now, in our variational principle we assume $\sigma $ to be independent of $%
g_{\mu \nu }.$ Variation of $\mathcal{L}_{DW}$ with respect to the canonical
variable $g^{\mu \nu }$ leads accordingly to%
\begin{equation}
\delta \mathcal{L}_{DW}=\frac{C}{a^{n-1}}\bar{A}^{\mu }\delta g_{\mu \tau
}+\delta \left( \frac{C}{a^{n-1}}\right) \bar{A}^{\mu }g_{\mu \tau }=\frac{C%
}{a^{n-1}}\bar{A}^{\mu }\left( \frac{1}{2}g_{\mu \tau }g_{\alpha \beta
}-g_{\mu \alpha }g_{\tau \beta }\right) \delta g^{\alpha \beta },
\end{equation}%
which, after substitution into (9), it implies%
\begin{equation}
S_{\alpha \beta }=-2\frac{\delta \mathcal{L}_{DW}}{\delta g^{\alpha \beta }}+%
\mathcal{L}_{DW}g_{\alpha \beta }=-2\frac{C}{a^{n-1}}\bar{A}^{\mu }\left( 
\frac{1}{2}g_{\mu \tau }g_{\alpha \beta }-g_{\mu \alpha }g_{\tau \beta
}\right) +\left( -\sigma +\mathcal{L}_{0}+\frac{C}{a^{n-1}}\bar{A}_{\tau
}\right) g_{\alpha \beta }.
\end{equation}%
One obtains%
\begin{equation}
S_{\tau \tau }=-\frac{C}{a^{n-1}}\bar{A}_{\tau }-\left( -\sigma +\mathcal{L}%
_{0}+\frac{C}{a^{n-1}}\bar{A}_{\tau }\right) =\left( -2\frac{C}{a^{n-1}}\bar{%
A}_{\tau }+\sigma -\mathcal{L}_{0}\right) 
\end{equation}%
or equivalently%
\begin{equation}
S_{\tau }^{\tau }=2\frac{C}{a^{n-1}}\bar{A}_{\tau }+\mathcal{L}_{0}-\sigma .
\end{equation}%
In the same manner one finds 
\begin{equation}
S_{\theta _{i}}^{\theta _{i}}=\mathcal{L}_{0}-\sigma .
\end{equation}

\textbf{Figure Caption:}

Fig. 1: The plot of radius $a\left( \tau \right) $ of the FRW universe for $%
n=4$, on the domain wall as a function of proper time. The oscillatory
behavior reveals a bounce at a distance greater than the horizon ($a>r_{h}$%
). The choice of parameters is : $C<0,$ $q=4.5,$ $m=6$ and $\ell =0.3.$ The
exact location of the event horizon ($r_{h}$) is shown in the smaller figure
for $f\left( r\right) .$


\begin{thebibliography}{9}
\bibitem{1} B. H Lee, W. Lee and M. Minamitsuji, Phys. Lett. B 679 (2009)
160.
\end{thebibliography}
\end{document}